\documentclass[12pt]{article}

\usepackage{newtxtext,newtxmath}
\usepackage{graphicx}
\usepackage{hyperref}
\usepackage[letterpaper,margin=1in]{geometry}

\renewenvironment{abstract}
	{\quotation}
	{\endquotation}

\date{}

\makeatletter
\renewcommand{\fnum@figure}{\textbf{Figure \thefigure}}
\renewcommand{\fnum@table}{\textbf{Table \thetable}}
\makeatother

\usepackage{scicite}

\usepackage{url}

\def\scititle{
	Validating LLMs in social science: Epistemic threats and emerging norms
}

\title{\bfseries \boldmath \scititle}

\author{
	Meera Desai$^{1\ast}$,
	Dallas Card$^{1}$,
	Abigail Z.\ Jacobs$^{1,2}$\and
	\small$^{1}$School of Information, University of Michigan, Ann Arbor, USA\and\
	\small$^{2}$Center for the Study of Complex Systems, University of Michigan, Ann Arbor, USA.\and
	\small$^\ast$Corresponding author. Email: madesai@umich.edu\and
}

\begin{document} 

\maketitle

\begin{abstract} \bfseries \boldmath
Large language models (LLMs) are reshaping social science methodology. Researchers increasingly prompt language models to generate quantitative measurements of social concepts, for example labeling data or simulating survey responses. Yet LLMs pose methodological challenges including bias, hallucination, and brittleness across contexts, with unclear threats to validity. Standard practices and norms for addressing these challenges are still emerging. We collect and systematically analyze validation practices in a comprehensive corpus of papers from eight flagship social science journals that use LLMs as measurement instruments. We find that LLM-generated measurements frequently play a central role in empirical analyses, yet validation practices are inconsistent and limited. We outline complementary strategies for more robust validation, pointing toward better norms and standards around the use of LLMs in social science. 
\end{abstract}

\section{Introduction}\label{sec:intro}
Across fields, social scientists are increasingly adopting the practice
of prompting large language models (LLMs) to generate quantitative
measures of social concepts. For example, researchers use LLMs as
measurement instruments by prompting LLMs to annotate and code data,
simulate survey responses, and estimate ideological positions. However,
the progression and legitimacy of social science rely on the ability to
make valid arguments. In turn, making valid arguments often rests on the
ability to take valid measurements of social constructs, like ideology,
emotion, or sentiment. This paper empirically examines the recent trend
of using LLMs as measurement instruments and identifies important
validity challenges (and opportunities) for social science research.

The widespread adoption of LLMs as measurement instruments reflects
their broad appeal as a faster, cheaper, and potentially more accurate
alternative to humans at tedious tasks like annotating or coding data
~\cite{gilardiChatGPTOutperformsCrowd2023, tornbergLargeLanguageModels2025}. Researchers also argue that LLMs are easier to use
than alternative computational approaches (such as training bespoke
machine learning models;
\cite{charnessNextGenerationExperimental2025}), and that their performance and flexibility makes them applicable
to a wide range of complex tasks than earlier computational text
analysis methods embraced by social scientists
~\cite{prussAIJurisprudenceLarge, rathjeGPTEffectiveTool2024a}.

Whether researchers use LLMs, dictionary-based approaches, human coding,
or any other measurement instrument, measuring abstract constructs
requires making interpretive choices about a concept's definition and
how it can be observed, leaving room for disagreement and error
~\cite{adcock2001measurement, goertzSocialScienceConcepts2006}. These methodological choices are consequential and
contestable, and validity cannot be assumed. Researchers have always had
to develop and engage with contextually appropriate ways of validating
their measurements even as computational tools have become increasingly
sophisticated ~\cite{changReadingTeaLeaves, quinn2010analyze, grimmerTextDataPromise2013}.

Despite the potential of LLMs as efficient tools for measuring social
concepts, there are plenty of reasons to be skeptical about using LLMs
as measurement instruments. Research from natural language processing
has highlighted many challenges associated with LLMs including their
sensitivity to small changes in prompt or configuration
~\cite{garcia-ferreroThisNotDataset2023, jangCanLargeLanguage, luFantasticallyOrderedPrompts2022, sclarQuantifyingLanguageModels2024, shuYouDontNeed2024, wangAreLargeLanguage2023, cummins2025threat}, bias
~\cite{feng-etal-2023-pretraining, santurkarWhoseOpinionsLanguage2023}, and poor calibration
~\cite{jiangHowCanWe, siReExaminingCalibrationCase2022, zhaoCalibrateUseImproving2021a}. Along with other model
limitations, these issues can affect the factual reliability of LLMs,
which sometimes produce ``hallucinated'' outputs that are coherent but
incorrect. Moreover, errors are difficult to predict or mitigate, as
neither the precise nature of model biases nor the relationship between
prompt and LLM performance are well understood
~\cite{messingHiddenMeasurementError2026, mittelstadtProtectScienceWe2023}. While methods for debiasing model
predictions using labeled data is an active area of research
~\cite{angelopoulosPredictionpoweredInference2023, egami2023using, ludwig2024large}, the improvements
from such methods tend to be small, and there is no guarantee they will
help in any particular case
~\cite{hullmanThisHumanStudy2026}.

Standard practices and norms for using LLMs as measurement instruments
are still emerging. While prior work investigates researchers' concerns
and motivations for using LLMs in social science research
~\cite{alveroGenerativeAISociological2025, liaoLLMsResearchTools2024, schroederLargeLanguageModels2025}, there has been little
investigation of current practices for using and validating LLMs as
measurement instruments.

In this paper, we collect a comprehensive corpus of papers where LLMs are used as measurement instruments across eight flagship social science journals. We analyze this corpus to
discover patterns in how LLMs
are being used in social science research and document and
explore approaches to validation. Drawing on the traditions of
measurement theory, we ask: \textbf{RQ1.} How do social scientists use
LLMs as measurement instruments in top social science journals?
\textbf{RQ2.} How do those researchers validate their measurements and
claims?

We find that LLM-generated measurements frequently play a central role
in empirical analyses, yet validation practices are limited and
inconsistent. We show how current practices are dominated by a single
aspect of construct validity, and outline complementary strategies for
more robust validation. Our work helps point the way forward for
development of better norms and standards around the use of LLMs for
social science.

\section{Results}\label{sec:findings}

\subsection{Overview and use cases}\label{overview-and-use-cases}
We collected 2,143 papers from top social science journals (see \hyperref[sec:methods]{Methods}) published between 2022 and late 2025.
Of these, we identified 50 measurement tasks in 27 papers that prompted an LLM to produce
quantitative measurements of a social concept, following the inclusion criteria in our codebook
(Average tasks per paper: 1.9, maximum 10.; see \hyperref[d.3-analytical-codes]{Additional materials and methods} and \hyperref[tab:journals]{Table S1}  for additional details). 
Note that these 27 articles are not a sampling of top social science papers using LLMs for measurement purposes. They are a comprehensive corpus, marking what is surely the first wave of what will become a much larger surge of research using these methods. In that way, these 27 articles serve as models for research to come, making it critical for us to understand the decisions these early researchers are making.

Most commonly, the papers in our corpus use LLMs to annotate or code
data (47 tasks, 25 papers), though two papers (3 tasks) used LLMs to
simulate survey participants. This distribution of task types is likely
influenced by our selection criteria, since we only consider uses of
LLMs that produce a quantitative measurement, such as a classification
or scaling, and only considered top social science journals.

Importantly, LLM-generated measurements often play a central role in
these papers. Most commonly, LLM-generated measurements serve as inputs
to the primary analysis, whereas in a smaller number of studies they are
used in a more limited capacity for data filtering or validation (e.g.,
generating reference labels for error analysis or serving as a
robustness check for another method) (see \hyperref[tab:purposes]{Table S2}). Given this centrality to research claims, rigorous reporting and
validation practices are essential, yet we identify limited and
inconsistent validation and reporting in practice.

The rest of this section is structured around the key decisions
researchers face when using LLMs as measurement instruments (drawing on
the measurement modeling framework from
\cite{adcock2001measurement, jacobsMeasurementFairness2021}): how they define the concepts they aim
to measure, how they design their instruments, and how they validate the
resulting measurements. We organize our results around three core
stages of the measurement process: conceptualization (\S 2.2), operationalization (\S 2.3), and validation (\S 2.4).

\subsection{Conceptualization: Without detailed definitions and validation, conceptual work is passed from researcher to model }\label{sec:findings-conceptualization}

Most social science concepts can be understood in multiple ways, and
some have highly contested definitions (e.g. ``terrorism,''
``privacy''), making it important to precisely define what one sets out
to measure \cite{bandalos2018measurement, goertzSocialScienceConcepts2006, krpanCallPrecisionStudy2025}. In practice, we find that
concepts were often underspecified in our corpus. Most commonly (13
papers, 29 tasks), researchers do not attempt to define the concept they
aim to measure, referring to it using only a single word or short
phrase.\footnote{We adapted a taxonomy of levels of concept
  specification in codebooks from
  \cite{haltermanWhatProtestAnyway2025}. See \hyperref[d.3-analytical-codes]{Additional materials and methods} in Supplementary Materials 
  for details.} For example,
\cite{rathjeGPTEffectiveTool2024a}
use the prompt ``\emph{Is the following post offensive? Answer only with
a number: 1 if offensive, and 0 if not offensive,''} without defining
``offensive.'' Detailed concept definitions that specify inclusion and
exclusion criteria are only included in prompts in a few cases (3
papers, 4 tasks). Even dictionary-style definitions, which provide more
high-level, generic characterizations (e.g., \emph{``An event is an
ongoing coherent situation.''}
\cite{rouhaniCollectiveEventsIndividual2023a}) are rarely included in prompts (7 papers, 7 tasks).

Even when researchers have carefully defined their constructs,
effectively incorporate these detailed operational definitions into
prompts is an active area of research: Halterman and Keith (2025)~\cite{haltermanCodebookLLMsAdapting2024} find
that including detailed stipulative definitions in prompts can sometimes
\emph{reduce} LLM performance, raising questions about how prompts are
interpreted. A small number of papers attempted to address this
empirically, iteratively refining their prompts against gold-standard
data such as human annotations, either through automated prompt
engineering (1 paper, 1 task) or informal trial-and-error (8 papers, 9
tasks). Beyond this, we found little discussion of how to effectively
incorporate conceptual definitions into prompts.

Overall, without a clearly defined concept, researchers using LLMs as
measurement instruments risk losing control of the concepts they
measure.

\subsection{Operationalization: Researchers using LLMs face many consequential experimental choices with little generalizable guidance }\label{sec:findings-operational}

Using an LLM in social science research requires making many design
choices including which prompt, model, and decoding/generation
parameters to use, and how to extract discrete answers from the LLM's
natural language response. Among reported design choices in our corpus
(\hyperref[tab:reporting]{Table S3}), we observe wide variation in the
specific configurations adopted in each of these measurement
instruments, which speaks to the wide range of choices researchers face
when using LLMs. This high degree of researcher freedom poses a
challenge for those using LLMs, as subtle changes in components can
greatly and unpredictably impact the performance and validity of the
instrument \cite{atrejaWhatsPromptLargeScale, sumanathilakaExploringImpactTemperature2025}. Yet
researchers\textquotesingle{} justifications for their component choices
often rely on intuition or examples that may not generalize, reflecting
the limited guidance currently available. Overall, our results
highlight that researchers using LLMs face many consequential
experimental choices with little guidance towards designing valid
LLM-based measurement instruments.

Among the papers in our corpus, prompting practices are highly diverse,
varying in length, structure, and content. For example, some prompts use
Markdown formatting or white space to separate different aspects of the
prompt. Though most studies in our corpus used zero-shot prompts, we
observed some instances of few-shot prompts.\footnote{Zero-shot prompts
  contain only task instructions and no (i.e., zero) labeled examples,
  whereas few-shot prompts include a small number of example
  input--output pairs to demonstrate the task.} Many prompts included
phrases attempting to constrain the model's output (e.g., ``Answer with a
number'' or ``Do not provide explanations'') and some prompts included
roles for the models (e.g., ``You are a helpful assistant'', ``You are a
medical expert diagnosing a patient.'').

Broadly, we found three main strategies for justifying prompt design:
researchers designed prompts by adhering to human annotation guidelines,
referencing existing literature (e.g., reusing prompts or strategies),
or testing a set of prompts empirically (e.g., ablations or limited
experiments). Each approach has significant limitations, which reflects
the lack of generalizable guidance\textbf{.} LLMs exhibit sensitivities
that can be difficult to predict
\cite{atrejaWhatsPromptLargeScale}, so it is unclear whether prompting strategies will generalize to
different tasks or data. Similarly, while ablation studies may aid
researchers in choosing a prompt from the infinite options, such studies
only identify the best prompt among a limited, arbitrarily chosen set of
variations reflecting the researcher's intuitions rather than
establishing any general superiority.

Another methodological choice researchers face is how to extract
quantitative answers (i.e. number or discrete category) from free-text
model outputs. Very few papers reported how they implemented this step
(4 papers, 6 tasks). Among those that did, some used string matching and
automated parsing methods (2 papers, 3 tasks). We also found one
paper/task that extracted labels by finding the maximum token
probability from a set of potential answer tokens, and another that used
human coders. 
The diversity of approaches and lack of reporting
highlights that while choice of answer extraction procedure could be
consequential for the resulting measurements, it is often overlooked or treated merely as a technical
implementation detail.

One potentially consequential yet underreported aspect of answer
extraction is how researchers handle cases where the model refuses to
produce an answer, often characterized as noncompliance
\cite{atrejaWhatsPromptLargeScale}. For example,
\cite{rathjeGPTEffectiveTool2024a}
described discarding data that triggered a content filter. In another
case, \cite{lemensUncoveringSemanticsConcepts2023} resampled the model repeatedly until it produced a numeric
response. It is often unclear what noncompliant responses indicate about
the data or how discarding them will impact the resulting measurements:
one paper\textquotesingle s analysis suggested that refused responses
reflected indeterminate or absent signals in the data
~\cite{lemensPositioningPoliticalTexts2025a} and another empirically demonstrated that removing refusals did
not impact their overall results
~\cite{bisbeeSyntheticReplacementsHuman2024}; however, these findings may not generalize across contexts. If
content that triggers filters or refusals is systematically different
from content that produces scoreable responses---for instance, if
certain sensitive topics are more likely to be filtered---these
procedures will introduce bias into downstream analyses. Without
reporting what data was discarded or replaced and why, researchers
cannot assess whether these decisions introduced systematic bias.

In addition to prompts and answer extraction procedures, we also
observed wide heterogeneity and inconsistent reporting in model choice
and decoding strategies. Additional details on the design choices and
justifications we found are provided in \hyperref[appendix-b-additional-results]{Supplementary text}.

Although researchers have embraced LLMs in part because of their
apparent ease of use, our findings illustrate the great number of
experimental decisions involved, and considerable variation in
instrument design across papers, often made seemingly arbitrarily. While
these choices do not inherently invalidate the measurements these
instruments take, they point to the need for proper documentation and
justification of these choices, and validation of the resulting
measurements.
\subsection{Validation}\label{sec:findings-validation}

Credible use of measurements of social concepts depends on evidence that
such measurements are valid. As discussed previously
(\S 2.3), using LLMs as measurement instruments entails making many
consequential experimental decisions with little guidance. The measurement framework proposes iteratively testing the \emph{construct validity} of the resulting measurements and updating these decisions accordingly. The construct validity framework offers researchers several complementary ways of assessing validity, which we enumerate in Figure \ref{fig:construct-valid-examples}.\footnote{Different social science traditions have different ontologies of ways to establish validity and reliability in research.  In some disciplines, construct validity is one type of validity, but it is common to consider an umbrella or unified definition of validity \cite{bandalos2018measurement, messick1989validity}. We describe one ontology under that umbrella from Jacobs \& Wallach (2021)~\cite{jacobsMeasurementFairness2021} and Wallach et al. (2025)~\cite{wallach2025position}, which draws heavily on Adcock \& Collier (2001)~\cite{adcock2001measurement}, Quinn et al. (2010)~\cite{quinn2010analyze}, and Messick (1996)~\cite{messick1996validity}, among others. This is not meant to be prescriptive, but should indicate what types of evidence are important for validity from LLM-based measurement instruments.} However, we find that researchers' validation practices among the papers in our corpus are limited and inconsistent. Moreover, we find that current practices are dominated by a single aspect of construct validity, (convergent
validity), and outline complementary strategies for more robust validation.

\subsubsection{ Tasks with missing or inconsistent validation }\label{sec:findings-no-val}

For 8 tasks across 6 papers, no efforts to validate LLM-generated
measurements were reported. These include tasks where the goal was to
filter data (1 paper, 1 task), to validate the main study (2 papers, 2
tasks), and to produce measurements for analysis in the paper's main
study (3 papers, 5 tasks). Interestingly, in some papers that contained
more than one measurement task using LLMs, researchers validated the
LLM-generated measurements for some but not all tasks. For example,  Sultan et al. (2024)~\cite{sultanSusceptibilityOnlineMisinformation2024a}
use LLMs to annotate headlines for four concepts, but only validate the
one concept most central to the paper's main question. This may suggest
that in some cases, not validating LLM-generated measurements may be
more a matter of priority, rather than lack of appropriate methods.

\subsubsection{Convergent validity is most common but practices are inconsistent}\label{sec:findings-converg}

Researchers in our corpus most commonly validate their LLM-generated
measurements by comparing them to gold-standard measurements of the same
construct (22 papers, 39 tasks), i.e., assessing \emph{convergent
validity.} However, we found considerable diversity in the
implementation of this common validation practice. This diversity
illustrates the corresponding breadth of methodological choices, each of
which shapes whether the comparison will actually provide meaningful
evidence of validity.

The quality of this gold-standard data impacts the soundness of the
validity assessment. In most cases, researchers used human-produced
gold-standard data (i.e., annotations or human survey data) (16 papers,
30 tasks). In these cases, the quality of human annotations hinges on
how they are collected, validated, and aggregated. Most studies in our
sample used multiple human annotators, in line with best practices ~\cite{boydstunQuantitativeContentAnalysis2023}. Annotator expertise varied considerably, ranging from
domain experts and trained reviewers to crowdworkers and students, and
was often not reported, despite evidence that annotator identity and
expertise matters
~\cite{plankProblemHumanLabel2022}.
Several studies also did not report intercoder reliability metrics for
their gold-standard datasets (5 papers, 6 tasks), making annotation
quality difficult to assess, while others reported only low or moderate
agreement, raising further questions about the suitability of these
annotations as a gold-standard reference.

In other cases (12 papers, 16 tasks) researchers compare LLM-generated
measurements to labels derived from other computational methods --- for
example, comparing LLM-generated annotations to annotations from a
fine-tuned classifier, LIWC, or dictionary methods 
--- instead or addition to using human labels. While sometimes useful,
these comparisons are only meaningful if the standard used for
comparison is itself valid. In our corpus, when researchers used an
off-the-shelf tool like LIWC to produce gold-standard labels (7 papers,
8 tasks), they mostly did not validate the performance of these tools on
their measurement task (5 papers, 6 tasks). By contrast, researchers who
developed or customized a tool (e.g. fine tuning an LLM classifier)
consistently validated its performance on their task (4 papers, 7
tasks). Without validation, the quality of the computationally produced
labels is unknown, and agreement with these labels provides little
meaningful evidence of validity.

The method of comparison between gold-standard datasets and
LLM-generated measurements also matters. Measurements are most often
compared using percent agreement or association metrics like Pearson
correlation values or Cramer's V. Yet, such metrics can inflate
perceived reliability because of not correcting for chance agreement and
masking potential sources of bias
~\cite{artsteinInterCoderAgreementComputational2008}. We also identify cases where the gold-standard human annotations
were drawn from extremely small samples of the full dataset (e.g., 50
out of 68,000 posts), raising questions about what the comparison can
tell us, particularly for diverse corpora or rare
categories.

\subsubsection{Other approaches to assessing construct validity }\label{sec:findings-non-converg}

\begin{figure}
    \centering
    \includegraphics[width=0.75\linewidth]{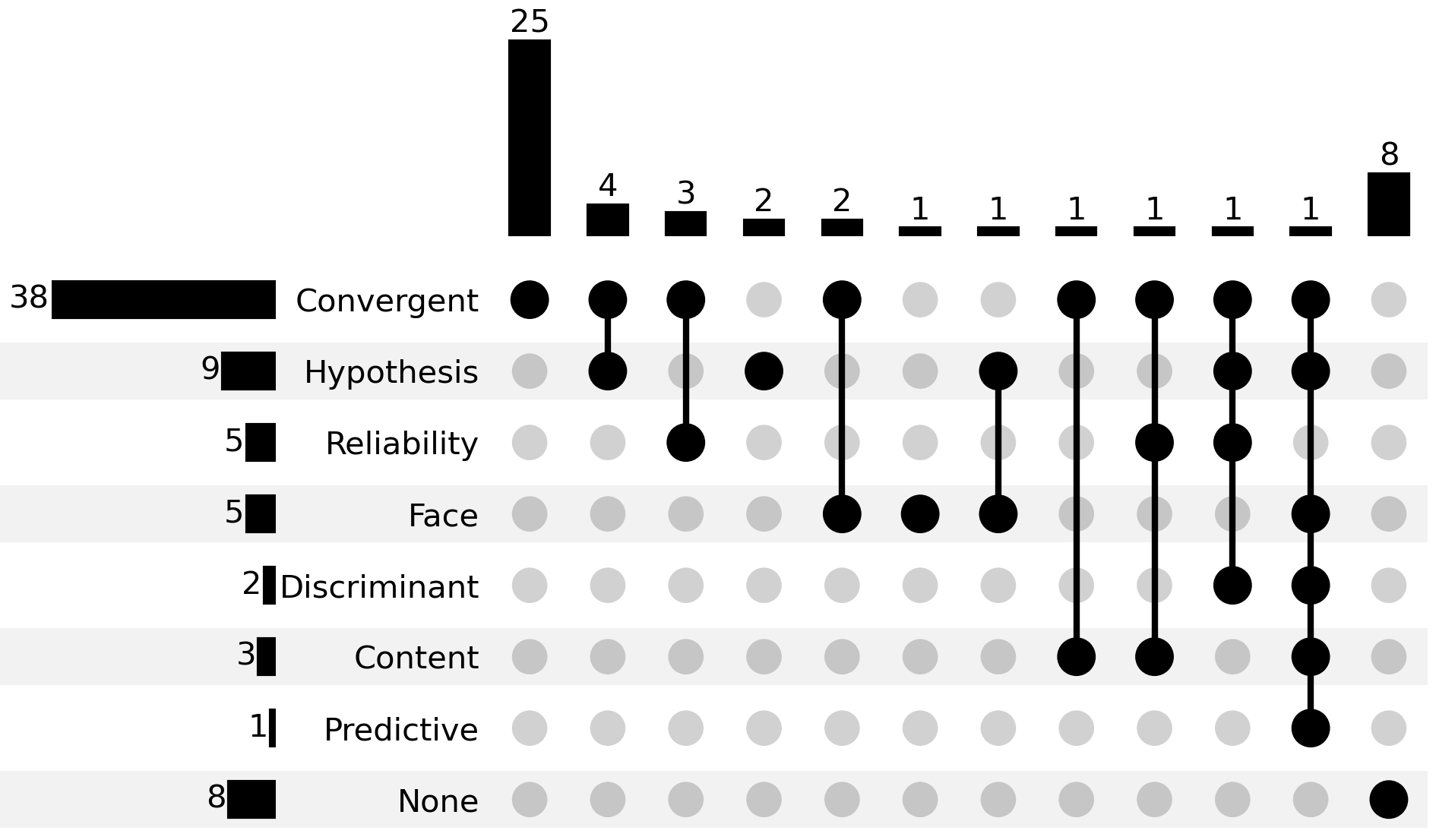}
    \caption{\textbf{Distribution of validation strategies across LLM-based measurement tasks.} 
UpSet plot showing which aspects of validity are assessed for each of the 50 measurement tasks in our corpus. Rows indicate aspects of construct validity and columns represent unique combinations of validation aspects. Convergent validity dominates (38 tasks) while other aspects of construct validity are rarely evaluated. Most tasks assess only a single validity type, and 8 tasks report no validation at all, highlighting narrow and uneven approaches to validation.
}
    \label{fig:val-distrib}
\end{figure}
Aside from convergent validity, researchers often overlook assessing
other aspects of validity (Figure \ref{fig:val-distrib}); indeed, validation is sometimes not reported or skipped
altogether. After convergent validity, we observe hypothesis validity
most frequently in our corpus, with a handful of papers assessing
multiple other aspects, including face validity and predictive validity.

\begin{figure}
    \centering
    \includegraphics[width=\linewidth]{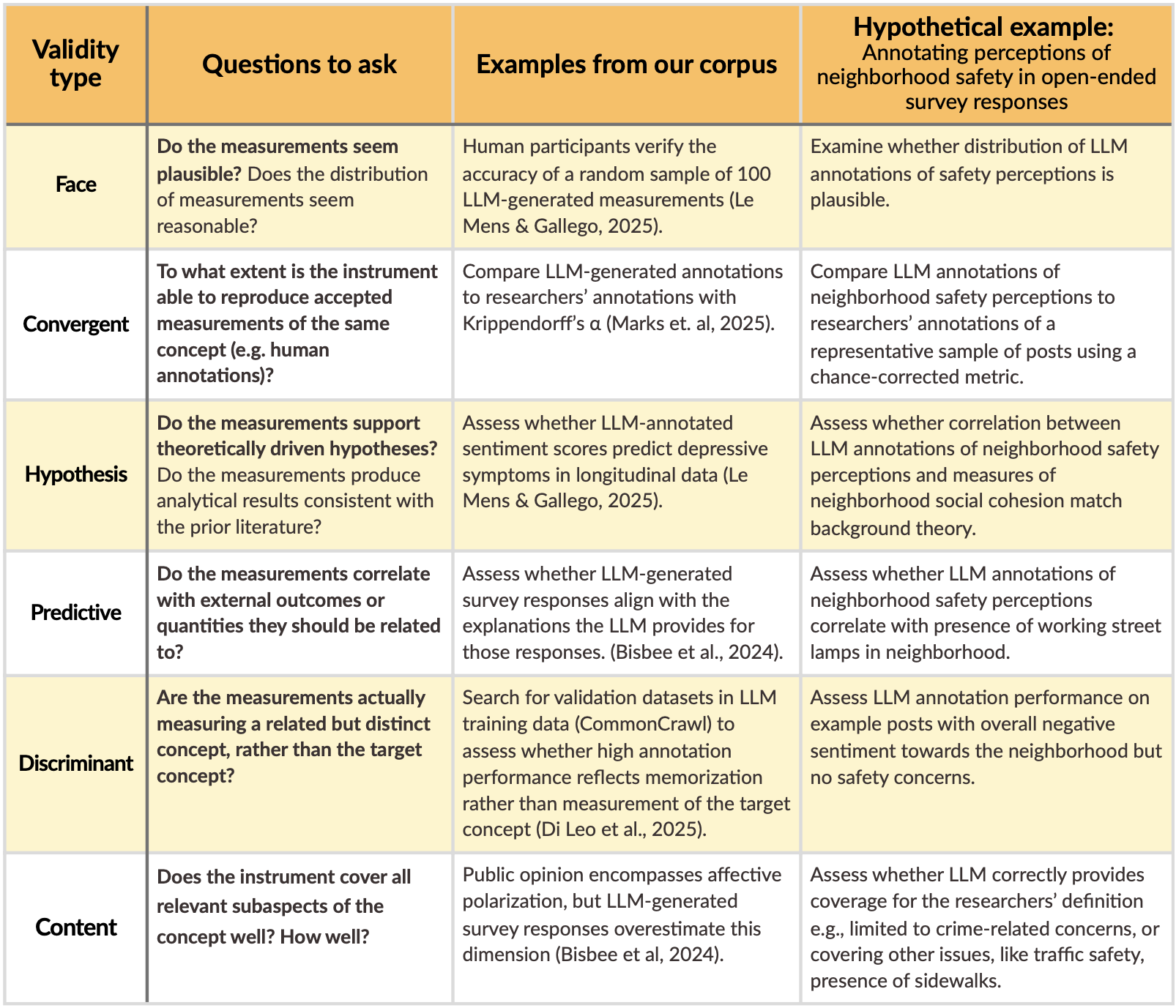}
    \caption{\textbf{Aspects of construct validity and their relevance to validating LLM-based measurement tasks, with examples.} We list guiding questions to elicit evidence of each aspect of validity, and provide a running hypothetical example. Together, we highlight the conceptual breadth of construct validity, and demonstrate how each aspect can be translated into concrete validation checks for LLM-based measurement tasks. For further discussion, see, e.g., Bandalos (2018)~\cite{bandalos2018measurement}, Goertz (2020)~\cite{goertz2020social}, Grimmer et al. (2012)~\cite{grimmer2012words}; for applications to generative AI, Wallach et al. (2025)~~\cite{wallach2025position}.
}
    \label{fig:construct-valid-examples}
\end{figure}

There are a range of ways to establish evidence for validity (recall
footnote 3). Yet in 28 out of 42 tasks where some attempt at validity is
reported, only one aspect of validity is assessed. Using more than one
aspect of validity offers different types of evidence that the
measurements are usefully capturing what they set out to measure. It is
notable that so few types of validity are typically invoked in these
papers; however, the diversity of approaches observed in our corpus
suggests that there is an actionable opportunity to develop more
thorough validation strategies for LLMs as measurement. We provide some
guiding examples in Figure \ref{fig:construct-valid-examples}  to help researchers reason about assessing various aspects of
validity when using LLMs.

\section{Discussion}\label{sec:discussion}
LLMs as measurement instruments offer a low barrier to large-scale
computational tools. However, when researchers prompt such models, they
are not applying some universal instrument to elicit meaningful
measurements. Instead, this practice relies on constructing a specific
measurement instrument with high degrees of researcher freedom and
(typically) low specificity about conceptualization, operationalization,
and validity. Because these measurement design choices can impact
measurement outcomes in ways that are difficult to predict or constrain,
thorough reporting and validation are critical for robust social science
research moving forward.

We observe a wide diversity of practices across instrument design,
reporting, and validation in published (i.e., normative) social science
research, which reflects the absence of established norms for this
emerging methodology. To support the development of such norms, we draw
on measurement theory and scholarship on social science methodology to
analyze existing practices and distill best practices from current
approaches: robustly validating LLM-generated measurements using
multiple lenses of construct validity, transparently reporting all
instrument components, and precisely defining the construct being
measured.

First, our findings highlight the importance of robustly validating
LLM-generated measurements, both in terms of the concept specification (\S \ref{sec:findings-conceptualization}) and model configuration
(\S\ref{sec:findings-operational}). In particular, when comparing LLM-generated measurements to
gold-standard data (most commonly human annotations), researchers should
follow best practices to ensure the gold-standard data is itself reliable. Best practices include using a formal codebook, training annotators for the task, collecting multiple annotations per example,
measuring annotator agreement using chance-corrected metrics, and
discussing disagreements to refine the concept. More broadly, we
recommend that researchers use multiple approaches to assessing
validity, informed by construct validity. Focusing solely on convergent
validity, i.e., comparing to a previous set of measurements, risks
laundering past poor conceptualizations and measurement choices into
future systems. We give some examples of how researchers might use
different lenses of validity in practice in Figure \ref{fig:construct-valid-examples}; beyond these examples, researchers should think creatively about
how each lens might apply to their measurement task and context.
Validating not just the measurements LLMs produce, but also the broader
constructs and arguments built upon them, is critical to achieve less
biased, more valid results.

Second, beyond validity, we find inconsistent and incomplete
methodological transparency. As we discuss in \S\ref{sec:findings-operational}, many studies omitted key components of their measurement
instruments, including exact prompt text, model version, decoding
functions, and procedures for extracting quantitative answers from
generated responses. In line with other efforts \cite{feuerriegel2026reporting}, we call for renewed attention to reporting these experimental details, especially as their relationship
to downstream bias and validity are under-studied.

Finally, researchers using LLM-generated measurements should make sure
to precisely define the concept they are measuring and record that
definition in their paper. When prompting LLMs to annotate or code data,
researchers should go through the process of hand annotating a subsample
of the data themselves: While the resulting annotations can be useful
for validation and bias correction, engaging with the annotation process
is itself independently valuable and encourages better theorizing
~\cite{santanaHowMachineLearning2024}. Working with annotators to process many examples, discuss
disagreements, and iteratively define a precise codebook is essential to
identifying the nature and boundaries of the construct, and should help
to produce documentation that will make the work clear and replicable.

Notably, we found underspecified concepts both in LLM prompts and
codebooks for human annotators. While precise definitions should be
documented in both cases, underspecified concepts may pose a greater
threat to validity for LLM-based measurement: Human annotators may share
a common understanding through disciplinary training, meetings among annotators, or may be used to
capture average human or expert interpretations (e.g., 
\cite{hoogheReliabilityValidity20022010}).
In contrast, when an LLM interprets an underspecified concept, there is
no such grounding: the model\textquotesingle s interpretation is neither
traceable to a particular scholarly tradition nor representative of any
identifiable human perspective, as some have argued
~\cite{ohaganMeasurementAgeLLMs2023, prussAIJurisprudenceLarge, wuLargeLanguageModels2023}. Instead, it reflects an
opaque aggregation of patterns in training data, making it impossible to
know whether the construct being measured aligns with the
researcher\textquotesingle s theoretical intent. Research suggesting
including detailed definitions in prompts can sometimes \emph{reduce}
LLM performance is further evidence that how models interpret concepts
and prompts is not well understood \cite{haltermanCodebookLLMsAdapting2024}.

Given this uncertainty, precise definitions do not always need to be
included in prompts. Instead, the primary purpose of precisely defining
a construct is to support validation, through which researchers can
verify that their measurements reflect the concept they set out to
measure. Without a precise notion of the concept one is measuring, it is
easy to conceptually drift from one implicit meaning to another,
effectively passing the conceptual control from the researcher to the
model ~\cite{haltermanSyntheticallyGeneratedText2025a, messeriArtificialIntelligenceIllusions2024}. This may carry particular
consequences for culturally variable concepts (e.g., ``political
extremism''), particularly as LLMs exhibit systematic cultural biases
~\cite{nadeemBiasBordersPolitical2026, santurkarWhoseOpinionsLanguage2023, sethHowDeepRepresentational2025}. Underspecified concepts
risk being measured through a particular cultural lens, potentially
leading to systematic misrepresentation of non-WEIRD populations.

Journals and reviewers have a role to play in establishing and enforcing
norms around all three of these priorities. Just as survey research is
expected to report questions and sampling procedures, research that
makes use of LLM-generated measurements should be held to equivalent
standards of disclosure. We encourage journals to develop explicit
reporting guidelines for LLM-based measurement that require disclosure
of the measurement instrument components alongside precise definition of
the construct being measured, justification of the conceptualizations
used in analysis, and validation strategies. Critically, such guidelines
should also require documentation of validation efforts including the
use of multiple lenses of validity. Editors and researchers might
reference exemplary work in this space for reference and inspiration
~\cite{bisbeeSyntheticReplacementsHuman2024, dileoMappingAIdeologyTaxonomy2024a, ohaganMeasurementAgeLLMs2023, wuLargeLanguageModels2023, zollerHumanAICollectives2025a}.

\section{Conclusion}\label{sec:conclusion}

Social science is positioned to be both revolutionized and undermined by
the rapid adoption of LLMs as these models become embedded in the
production of quantitative measures. Our analysis shows that
LLM-generated measurements are already central to many empirical claims
in top journals, yet the practices used to design and validate these
instruments remain inconsistent and limited. Because LLM-based measures
are the product of numerous consequential and weakly theorized design
choices, making credible use of these measurements depends on precise
conceptualization, transparent reporting, and rigorous evaluation of
construct validity. By documenting current practices and clarifying the
stakes, we highlight a growing gap between the centrality of
LLM-generated measurements in empirical analysis and the maturity of the
validation practices used to justify them. The development and
institutional adoption of stronger reporting and validating norms will
be essential to ensuring that the adoption of LLMs contributes to the
robust progress in social science research.

\section{Materials and Methods}\label{sec:methods}

\subsection{Data collection}\label{data-collection}

To assess how LLMs are being used in high caliber social science
research, we compiled a corpus of recently published papers that used
LLMs to generate quantitative measurements of social concepts. For
coverage across fields, we selected top journals in political science,
sociology, psychology, and the multidisciplinary journal \emph{Nature
Human Behaviour}. Within political science, we selected the top two
journals according to
\textit{SCImago
Journal Rank} \cite{scimago_polisci}. Since prompting LLMs for quantitative measurement is a
relatively recent methodological innovation, we also added a top
methodology-focused political science journal (\emph{Political
Analysis}) where such cutting edge methodologies are typically explored.
For psychology, we collected papers from \emph{PNAS} under the
Psychological
and Cognitive Sciences topic. We also searched top sociology journals,
but did not find any papers involving prompting LLMs in \emph{American
Journal of Sociology, American Sociological Review}, or the \emph{Annual
Review of Sociology}. Finally, we included articles from \emph{Nature
Human Behaviour} to capture interdisciplinary social science research
using this emerging methodology.

We downloaded all research articles with appendices published in these
journals from 2022 to September 2025. From this set, we selected the
articles that included one of the following keywords anywhere in the
article or its appendix: ``LLM'', ``AI '', ``GPT'', ``language model'', ``Llama'',
or ``Claude.'' This list of keywords was reviewed by all authors. The
resulting subset of papers was further screened by two coders for
inclusion if the paper included prompting an LLM to produce a numerical
or categorical measure of a social concept.

Because our study focuses on social science applications rather than
model evaluation, we generally excluded papers in which the task
primarily focused on measuring the model's internal structure,
capabilities, or behavior (e.g. measuring LLMs ability to deceive or
implicit bias in LLMs). However, we included a subset of such papers
when they included prompting tasks designed to explicitly measure a
social concept external to the model. For example, Rathje et al.,
(2024), ask whether GPT is an effective tool for multilingual
psychological text analysis, i.e., a research question about model
capabilities. We included this paper, however, because part of their
data collection involves using LLMs to measure sentiment in tweets, a
social science concept rather than a property of models. See our
codebook 
for further examples and details (\hyperref[d.3-analytical-codes]{Additional materials and methods}).

None of the papers from 2022 met these criteria, so after screening, the
final set of papers was published between 2023 and 2025. Some papers
included more than one task that met our inclusion criteria, so in our
results we distinguish between \emph{papers} and \emph{tasks}.

\subsection{Analysis}\label{analysis}

One author served as the primary coder, supported by a graduate research
assistant. Working through separate portions of the corpus (each
approximately half of the corpus), both coders applied descriptive codes
capturing the components of each prompting task, including the prompt,
the model used, the justifications for each component, and the overall
validation process reported in the paper. The primary coder then
reviewed the research assistant's coding, discussing disagreements and
resolving ambiguities. To develop analytic codes, all authors discussed
emerging themes in the descriptive codes. Drawing on these discussions
and relevant literature from measurement theory and quantitative content
analysis methodology, the primary coder developed analytic codes to
capture higher-order dimensions of each task, including how the target
concept was conceptualized in the prompt and which aspects of construct
validity each paper used. When cases arose that were unclear or
challenged the codebook, the primary coder brought these to the full
author team for discussion and updated the codebook accordingly. Our
codebook is available  
in \hyperref[d.3-analytical-codes]{Additional materials and methods} in Supplementary Materials.

We release our qualitative coding results as a dataset to accompany this paper here: \\ \texttt{\detokenize{
https://osf.io/ab5zc/overview?view_only=f8e3fe5a36e0415aa4c441ad061e8ccc}}. This includes the titles and DOIs of all papers considered and selected for our corpus, and qualitative codes across 33 dimensions of instrument design and validation.

\paragraph{Acknowledgments}
Abigail Jacobs and Meera Desai were supported in part by the
Microsoft Research AI \& Society Fellowship. We are grateful to  Seorin Jang for his help with data selection, and Chloe Yueh for her assistance in qualitative coding. We would also like to thank Amber Boydstun, Andrew Halterman, Jeffrey Lockhart, and Michael Thompson-Brusstar for their helpful feedback and comments. 

\nocite{argyleOutOneMany2023a}

\bibliography{science_template}
\bibliographystyle{sciencemag}

\newpage
\renewcommand{\thefigure}{S\arabic{figure}}
\renewcommand{\thetable}{S\arabic{table}}
\renewcommand{\theequation}{S\arabic{equation}}
\renewcommand{\thepage}{S\arabic{page}}
\setcounter{figure}{0}
\setcounter{table}{0}
\setcounter{equation}{0}
\setcounter{page}{1}
\begin{center}
\section*{Supplementary Materials for\\ \scititle} \label{appendix}

\author{

	Meera Desai$^{1\ast}$,
	Dallas Card$^{1}$,
	Abigail Z.\ Jacobs$^{1,2}$

	\small$^{1}$School of Information, University of Michigan, Ann Arbor, USA\and\
	
    \small$^{2}$Center for the Study of Complex Systems, University of Michigan, Ann Arbor, USA.\and

    \small$^\ast$Corresponding author. Email: madesai@umich.edu\and

}
\end{center}

\subsubsection*{This PDF file includes:}
Additional materials and methods \\
Supplementary text\\
Tables S1 to S3\\

\subsection*{Additional materials and methods}\label{appendix-d---codebook}
\subsubsection*{Inclusion criteria
}\label{d.1-inclusion-criteria}

High-level selection criteria: Prompts a large language model to
generate a category (from a set list of categories) or a number of a
social science concept

\begin{itemize}
\item
  Exclude studies that use LLMs without prompting (i.e. BERT classifier)

  \begin{itemize}
  \item
    Exclude studies that use LLMs to produce or interpret embeddings
  \end{itemize}
\item
  Exclude studies that do not measure an abstract social concept. Social
  concepts are attitudes (e.g., emotion), identity (e.g., gender
  identity), behavior (e.g., expressed support of a topic), or
  attributes of social phenomena (e.g., essay quality) that are defined
  relative to people in social contexts.

  \begin{itemize}
  \item
    Positive example: sentiment
  \item
    Negative example: distance
  \end{itemize}
\item
  Exclude studies that are not social science (i.e., there are some
  neuroscience studies among the PNAS papers). Social science is the study
  of human behavior, interactions, and societies, exploring how people
  live, work, and govern themselves. Studies that focus only on
  biological substrates like neural activity, connectivity, physiology
  should be excluded.

  \begin{itemize}
  \item
    Edge case: However, studies that use LLMs to produce diagnoses from
    text should be included, as diagnosis in this case is an act of
    interpretation of socially produced language.
  \end{itemize}
\item
  Exclude studies where the goal is to evaluate model capabilities,
  internal structure, or behavior (e.g., measuring an
  LLM\textquotesingle s ability to deceive or its implicit bias).
  However, include studies that contain subtasks where LLMs are used to
  measure a social concept independent of the model itself

  \begin{itemize}
  \item
    Positive example: Rathje et al., (2024), a study asking whether GPT
    is effective for multilingual psychological text analysis, would be
    included because it involves using LLMs to measure sentiment in
    tweets, and sentiment is a social concept independent of the model
  \item
    Negative example: Cheung et al., (2025)~\cite{cheungLargeLanguageModels2025a}, evaluates whether LLMs exhibit cognitive biases in moral
    decision making. By having LLMs take multiple choice moral
    decision-making tests, the authors do use LLMs to produce
    quantitative measures of a social phenomena. However, the moral
    decision-making being measured is a property of the model, not a
    social concept independent of it --- the study tells us something
    about LLM behavior, not about human behavior or social phenomena.
  \end{itemize}
\end{itemize}

\subsubsection*{Descriptive codes}\label{d.2-descriptive-codes}
\begin{description}

\item[\normalfont\textbf{Prompt.}] The actual text of the prompt used. If none, write ``not reported.'' If the exact text is not given, select the sentence in the paper that describes the prompt. \textit{Example: ``Please classify the following news media social media post as either negative for Republicans (Democrats) (1) or not negative for Republicans (Democrats) (0).''}

\item[\normalfont\textbf{Prompt justification.}] Any justification for the selection and/or design of the prompt, including: descriptions of trying alternative prompts; citations of papers that suggest certain prompt structures; reasoning for why the prompt has a certain structure. If none, write ``not reported.'' \textit{Example: ``For all template fragments, phrasing was selected to closely match the ANES, although the ANES phrasing was translated into first-person declarations.''}

\item[\normalfont\textbf{Concept.}] The concept being measured in the task, including: the systematization or detailed definition of the concept; how the concept is operationalized (e.g., continuous spectrum collapsed to binary label); justification for a particular definition; discussions of the concept's contestedness. \textit{Example: ``We defined impulsiveness as the `likelihood for a spontaneous purchase and instant gratification potential.'''}

\item[\normalfont\textbf{Model.}] The specific LLM(s) used, in the most detail provided (i.e., version numbers, API access point, etc.). If none, write ``not reported.'' \textit{Example: ``GPT-4o (gpt-4o-2024-05-13), GPT-4 Turbo (gpt-4-turbo-2024-04-09), GPT-4o mini (gpt-4o-mini-2024-07-18) and GPT 3.5 Turbo (gpt-3.5-turbo-0125).''}

\item[\normalfont\textbf{Model justification / selection.}] Any justification for model selection, including: descriptions of trying alternative LLMs; citations from research papers justifying model selection; arguments about performance or reliability; model characteristics that drove selection (e.g., instruction-tuned, value-aligned); discussions of open vs.\ closed models or cost. If none, write ``not reported.'' \textit{Example: ``Four are high-performing closed-sourced models\ldots The other four are open-sourced Llama-based models.''}

\item[\normalfont\textbf{Decoding strategy.}] Temperature setting, top-$p$ or top-$k$ value, greedy decoding, or mention of using default settings. If none, write ``not reported.'' \textit{Example: ``We opted for the default setting of 1 to allow for reasonable variations, which best represents the learned probability distribution.''}

\item[\normalfont\textbf{Decoding justification.}] Justification for the decoding strategy selection, including descriptions of other strategies considered or tested. If none, write ``not reported.'' \textit{Example: ``We set the temperature parameter to 0, to ensure that the LLM would generate its response by selecting the most likely next token, and thus make the LLM responses as deterministic as possible.''}

\item[\normalfont\textbf{Answer extraction.}] Description of how categorical or numerical answers are extracted from model responses, or how prompts are formulated to maximize answer extraction (e.g., prompting to produce JSON). If none, write ``not reported.'' \textit{Example: ``Whenever this option was available, we set the response format to be a JSON object.''}

\item[\normalfont\textbf{Dealing with non-compliant answers.}] Description of how model responses that do not produce a usable category or label are standardized, or discussion of this problem. If none, write ``not reported.'' \textit{Example: ``Even when explicitly instructed to provide answers in a specific format, some LLMs did not always comply and occasionally returned verbose responses\ldots We therefore removed the response until the first line break if the response started with `Sure,\ldots' ''}

\item[\normalfont\textbf{Validation.}] Anything done to ensure the variables generated by the instrument are valid, including: comparison to human-generated variables or other computational methods; comparison of downstream analyses across methods; justification and discussion of validation practices. Code only for validation of the prompt-based instrument's output. If none, write ``not reported.'' \textit{Example: ``We also found that the sentiment scores generated by ChatGPT for each individual were highly correlated with the human sentiment scores ($\rho = 0.96$, $p < 0.001$).''}

\item[\normalfont\textbf{Validation: human-generated variable procedure.}] If compared to human-generated variables, description of the annotation process (e.g., annotation procedure, annotator expertise, validation of annotations). \textit{Example: ``We recruited human raters ($N = 470$) and asked them to rate a random subset of the collected written responses in terms of their affective tone.''}

\item[\normalfont\textbf{Validation: other computational method.}] If compared to another computational method, description of that method. \textit{Example: ``We also trained a BERT-based supervised probabilistic text classifier using the crowdworkers' ratings collected by Benoit et al.\ (2016).''}

\item[\normalfont\textbf{Other model parameters.}] Descriptions of any other parameters not covered by other codes, including: hardware setup, random seeds, max tokens, etc. \textit{Example: ``\texttt{max\_tokens: 20}. This parameter cuts the response of the LLM to a maximum of 20 tokens.''}

\item[\normalfont\textbf{Downstream use of LLM-generated variable.}] Description of how the variables generated by the instrument will be used in downstream analysis, ideally including: (a) the research question the variables address and (b) the method used to answer it. May also include how variables are processed for downstream analysis. If none, write ``not reported.'' \textit{Example: ``Robust linear regression predicted future depressive symptom scores (PHQ-9) after three weeks using ChatGPT (GPT-4) sentiment ratings.''}

\item[\normalfont\textbf{Number of samples.}] Number of times the task is performed (e.g., for a data annotation task, how many items are annotated).

\end{description}

\subsubsection*{Analytical codes\\
}\label{d.3-analytical-codes}

\noindent Analytical codes:

Conceptualization:

\begin{itemize}
\item
  Concept in prompt: Indicates how the specific concept is defined in
  the prompt or study. This code captures the level of specificity used
  to explain the concept being examined.\\
  OPTIONS (adapted from
  \cite{haltermanWhatProtestAnyway2025}):

  \begin{itemize}
  \item
    Single word or short phrase: The concept is mentioned briefly
    without further explanation (e.g., ``fairness,'' ``bias'').

    \begin{itemize}
    \item
      Example: \emph{``Is the following (Turkish) post offensive?''}
    \end{itemize}
  \item
    Dictionary definition: The prompt only defines the concept using a
    standard or generic definition.

    \begin{itemize}
    \item
      Example: \emph{``Is the following text focused on the duty to vote
      as independent (voting as a duty to oneself, to make
      one\textquotesingle s voice heard, to express opinions, to
      exercise rights, to take action, etc.) or as interdependent
      (voting as a duty to others, to community, to children, to
      history)?''}
    \end{itemize}
  \item
    Stipulative definition - The prompt includes a detailed definition
    of the concept with inclusion and exclusion criteria.

    \begin{itemize}
    \item
      Example: ``\emph{You are a helpful assistant tasked with labeling
      whether a social media post from an American Democrat or
      Republican expresses solidarity with, or positive emotions
      towards, the poster's party, including specific party members, the
      whole party, or political allies in general. You should label
      posts as expressing ingroup solidarity with the poster's party
      only if they describe solidarity with or amongst party members,
      indicate liking of or pride in the poster's political allies,
      and/or mention the unity or strength or competency of the poster's
      political allies or party. This includes all posts where the
      poster praises the achievements or views of their party, talks
      about party members collaborating or supporting one another, or
      framing the poster's party as good, competent, popular, strong or
      moral people. A post expresses ingroup solidarity if it is
      directed at political allies, not if it is directed at political
      opponents or apolitical people. Take a moment to think, but only
      answer with `yes' if the post expresses ingroup solidarity, or
      `no' if it does not express ingroup solidarity. Label this
      post:''}
    \end{itemize}
  \end{itemize}
\end{itemize}

\noindent Validation:

\begin{itemize}
\item
  Validation aspects: The type of validity used to validate the
  measurement of the concept. (Drawn from
  \cite{jacobsMeasurementFairness2021})\\
  OPTIONS

  \begin{itemize}
  \item
    Face: The reasonableness of the LLM-generated measurements is
    assessed through a quick test.

    \begin{itemize}
    \item
      Validity is assessed through informal, non-systematic sampling of
      model outputs or explanations to check whether the measurement
      procedure appears reasonable
    \item
      Examples: subset of outputs is reported and/or examined manually
      for reasonableness, reasonableness of distribution of outputs is
      assessed
    \end{itemize}
  \item
    Convergent: Validation is performed by assessing the agreement or
    correlation with another measure of the same concept where both
    measures are intended to operationalize the same concept rather than
    predict an external outcome (the latter is predictive validity).

    \begin{itemize}
    \item
      Convergent validity can be assessed for the concept the LLM
      directly measures or the downstream concepts those measurements
      are used to measure
    \item
      Examples:

      \begin{itemize}
      \item
        LLM-generated ideological positions of tweets are compared to
        human annotations of ideological positions of same tweets
      \item
        LLM-generated ideological positions of tweets are used to
        measure ideological positions of \emph{parties}, and these
        estimates are compared to estimates measured using human
        annotations
      \end{itemize}
    \end{itemize}
  \item
    Hypothesis: The ability of the LLM-generated measurements to answer
    a theoretically interesting question is assessed.

    \begin{itemize}
    \item
      The question must be theoretically grounded, not just empirical
    \item
      A null or negative result still counts as assessing hypothesis
      validity
    \item
      Examples: see whether estimates of party ideology from
      GPT-generated pair rankings can be used to estimate change in
      party positions over time
    \end{itemize}
  \item
    Content: The extent to which the LLM-generated measurements capture
    the full spectrum of the concept is assessed.

    \begin{itemize}
    \item
      LLM performance is assessed on sub-dimensions of the target
      concept, often done using error analysis. However, the error
      analysis must be related to the concept of interest, and the
      relationship should be discussed.

      \begin{itemize}
      \item
        Example: Public opinion encompasses affective polarization, but
        LLM-generated survey responses overestimate this dimension.
      \end{itemize}
    \item
      If measurements are also validated using convergent validity,
      distribution of validation data is discussed and/or accounted for

      \begin{itemize}
      \item
        Example: Note that accuracy on validation data is mostly driven
        by negative cases as data is skewed towards negative examples
      \end{itemize}
    \item
      Example: Find that LLM performance holds for low-frequency
      categories, which demonstrates that the measurements cover the
      full distribution of the concept rather than only its most common
      manifestation.
    \end{itemize}
  \item
    Predictive: The extent to which the LLM-generated measurements
    correlate as expected with related but external quantities is
    assessed.

    \begin{itemize}
    \item
      Quantity must not be theoretically related to LLM-generated
      measurements (that is hypothesis validity)
    \item
      Example: Prompt LLM to generate explanations for each measurement
      and find that these explanations correlate with LLM-generated
      measurements.
    \end{itemize}
  \item
    Discriminant: The extent to which the LLM-generated measurements
    measure a concept other than the target concept is assessed.

    \begin{itemize}
    \item
      Most likely assessed through correlations between LLM generated
      measurements and unrelated concepts.
    \item
      Example:
    \end{itemize}
  \end{itemize}
\item
  For human annotations (drawn partially from
  Boydstun (2023)~\cite{boydstunQuantitativeContentAnalysis2023}):

  \begin{itemize}
  \item
    Used human data for validation: whether human-generated data was
    used in validating the measurement of concept.\\
    OPTIONS:

    \begin{itemize}
    \item
      Yes (annotations): Human annotators labeled or evaluated data for
      validation.
    \item
      Yes (not annotations): Human data in another form was used (e.g.,
      survey responses, behavioral data).
    \item
      No: No human data was used in the validation process.
    \end{itemize}
  \item
    If yes: Collected own human data: whether the authors collected new
    human data themselves for validation.\\
    OPTIONS:

    \begin{itemize}
    \item
      Yes: The study collected original human data (e.g., Authors
      collected annotations by labeling items themselves or oversaw a
      data collection process where people other than the authors
      collected the data)
    \item
      No: Authors reused previously published data or compiled (combined
      or selected elements from) existing datasets
    \item
      Not applicable: No human data was collected.
    \end{itemize}
  \item
    Annotator expertise: The type of individuals who performed the
    annotations.\\
    OPTIONS:

    \begin{itemize}
    \item
      Crowdworkers: Annotators recruited through crowdsourcing platforms
    \item
      Experts: Domain specialists with formal expertise.
    \item
      Students: Student annotators, who are often recruited from a
      university.
    \item
      Participants: Study participants who are not necessarily trained
      annotators.
    \item
      Authors: The researchers themselves performed the annotations.
    \item
      Not applicable: human data was not used.
    \end{itemize}
  \item
    Inner annotator agreement reported or disagreements resolved:
    whether the study reported agreement between annotators or described
    how disagreements were resolved.\\
    OPTIONS:

    \begin{itemize}
    \item
      Agreements reported: The study reports inter-annotator agreement
      metrics (e.g., Cohen's kappa).
    \item
      Disagreements resolved through discussion: Annotators discussed
      and reconciled disagreements.
    \item
      Disagreement resolved through majority vote: Final labels
      determined by majority agreement among annotators.
    \item
      No reporting on disagreement resolution or agreement: The study
      does not describe agreement or resolution methods.
    \item
      Not applicable {[}one annotator{]}: Only one annotator performed
      the labeling.
    \item
      Other {[}validated in another way{]}: Validation occurred through
      alternative methods.
    \end{itemize}
  \end{itemize}
\end{itemize}

\subsubsection*{List of publications included in our
analysis}\label{appendix-c---list-of-publications-included-in-our-analysis}

\begin{quote}
Argyle, L. P., Busby, E. C., Fulda, N., Gubler, J. R., Rytting, C., \& Wingate, D. (2023). Out of One, Many: Using Language Models to Simulate Human Samples. Political Analysis, 31(3), 337--351. https://doi.org/10.1017/pan.2023.2
\\[0.5\baselineskip]
Bayerl, A., Dover, Y., Riemer, H., \& Shapira, D. (2024). Gender rating gap in online reviews. Nature Human Behaviour, 9(3), 507--520. https://doi.org/10.1038/s41562-024-02003-6
\\[0.5\baselineskip]
Bhatia, S., Van Baal, S. T., Wang, F., \& Walasek, L. (2025). Computational analysis of 100 K choice dilemmas: Decision attributes, trade-off structures, and model-based prediction. Proceedings of the National Academy of Sciences, 122(17), e2406489122. https://doi.org/10.1073/pnas.2406489122
\\[0.5\baselineskip]
Bisbee, J., Clinton, J. D., Dorff, C., Kenkel, B., \& Larson, J. M. (2024). Synthetic Replacements for Human Survey Data? The Perils of Large Language Models. Political Analysis, 32(4), 401--416. https://doi.org/10.1017/pan.2024.5
\\[0.5\baselineskip]
Di Leo, R., Zeng, C., Dinas, E., \& Tamtam, R. (2024). Mapping (A)Ideology: A Taxonomy of European Parties Using Generative LLMs as Zero-Shot Learners. SSRN. https://doi.org/10.2139/ssrn.4907347
\\[0.5\baselineskip]
Evans, D., Mason, C., Chen, H., \& Reeson, A. (2024). Accelerated demand for interpersonal skills in the Australian post-pandemic labour market. Nature Human Behaviour, 8(1), 32--42. https://doi.org/10.1038/s41562-023-01788-2
\\[0.5\baselineskip]
Garg, P., \& Fetzer, T. (2025). Political expression of academics on Twitter. Nature Human Behaviour, 9(9), 1815--1832. https://doi.org/10.1038/s41562-025-02199-1
\\[0.5\baselineskip]
Goergen, J., De Bellis, E., \& Klesse, A.-K. (2025). AI assessment changes human behavior. Proceedings of the National Academy of Sciences, 122(25), e2425439122. https://doi.org/10.1073/pnas.2425439122
\\[0.5\baselineskip]
Heseltine, M., Barnehl, H., \& Wojcieszak, M. (2025). Partisan temporal selective news avoidance: Evidence from online trace data. American Journal of Political Science, 69(4), 1541--1558. https://doi.org/10.1111/ajps.12944
\\[0.5\baselineskip]
Hulme, M. P. (2025). War and Responsibility. American Political Science Review, 1--24. https://doi.org/10.1017/S0003055425000206
\\[0.5\baselineskip]
Hur, J. K., Heffner, J., Feng, G. W., Joormann, J., \& Rutledge, R. B. (2024). Language sentiment predicts changes in depressive symptoms. Proceedings of the National Academy of Sciences, 121(39), e2321321121. https://doi.org/10.1073/pnas.2321321121
\\[0.5\baselineskip]
Le Mens, G., \& Gallego, A. (2025). Positioning Political Texts with Large Language Models by Asking and Averaging. Political Analysis, 33(3), 274--282.\\ https://doi.org/10.1017/pan.2024.29
\\[0.5\baselineskip]
Le Mens, G., Kovács, B., Hannan, M. T., \& Pros, G. (2023). Uncovering the semantics of concepts using GPT-4. Proceedings of the National Academy of Sciences, 120(49), e2309350120. https://doi.org/10.1073/pnas.2309350120
\\[0.5\baselineskip]
Lee, B., Aiyappa, R., Ahn, Y.-Y., Kwak, H., \& An, J. (2025). A semantic embedding space based on large language models for modelling human beliefs. Nature Human Behaviour, 9(9), 1928--1940. https://doi.org/10.1038/s41562-025-02228-z
\\[0.5\baselineskip]
Lehr, S. A., Saichandran, K. S., Harmon-Jones, E., Vitali, N., \& Banaji, M. R. (2025). Kernels of selfhood: GPT-4o shows humanlike patterns of cognitive dissonance moderated by free choice. Proceedings of the National Academy of Sciences, 122(20), e2501823122. https://doi.org/10.1073/pnas.2501823122
\\[0.5\baselineskip]
Marks, M., Kyrychenko, Y., Gärdebo, J., \& Roozenbeek, J. (2025). Ingroup solidarity drives social media engagement after political crises. Proceedings of the National Academy of Sciences, 122(35), e2512765122. https://doi.org/10.1073/pnas.2512765122
\\[0.5\baselineskip]
Park, J. S., Gollapudi, K., Ke, J., Nau, M., Pappas, I., \& Leong, Y. C. (2025). Emotional arousal enhances narrative memories through functional integration of large-scale brain networks. Neuroscience. https://doi.org/10.1101/2025.03.13.643125
\\[0.5\baselineskip]
Rathje, S., Mirea, D.-M., Sucholutsky, I., Marjieh, R., Robertson, C. E., \& Van Bavel, J. J. (2024). GPT is an effective tool for multilingual psychological text analysis. Proceedings of the National Academy of Sciences, 121(34), e2308950121.\\https://doi.org/10.1073/pnas.2308950121
\\[0.5\baselineskip]
Rouhani, N., Stanley, D., COVID-Dynamic Team, Adolphs, R., . (2023). Collective events and individual affect shape autobiographical memory. Proceedings of the National Academy of Sciences, 120(29), e2221919120. \\https://doi.org/10.1073/pnas.2221919120
\\[0.5\baselineskip]
Salvi, F., Horta Ribeiro, M., Gallotti, R., \& West, R. (2025). On the conversational persuasiveness of GPT-4. Nature Human Behaviour, 9(8), 1645--1653. \\https://doi.org/10.1038/s41562-025-02194-6
\\[0.5\baselineskip]
Sultan, M., Tump, A. N., Ehmann, N., Lorenz-Spreen, P., Hertwig, R., Gollwitzer, A., \& Kurvers, R. H. J. M. (2024). Susceptibility to online misinformation: A systematic meta-analysis of demographic and psychological factors. Proceedings of the National Academy of Sciences, 121(47), e2409329121. https://doi.org/10.1073/pnas.2409329121
\\[0.5\baselineskip]
Velez, Y. R., \& Liu, P. (2025). Confronting Core Issues: A Critical Assessment of Attitude Polarization Using Tailored Experiments. American Political Science Review, 119(2), 1036--1053. https://doi.org/10.1017/S0003055424000819
\\[0.5\baselineskip]
Viganò, S., Bayramova, R., Doeller, C. F., \& Bottini, R. (2024). Spontaneous eye movements reflect the representational geometries of conceptual spaces. Proceedings of the National Academy of Sciences, 121(17), e2403858121. \\https://doi.org/10.1073/pnas.2403858121
\\[0.5\baselineskip]
Waldfogel, H. B., Dittmann, A. G., \& Birnbaum, H. J. (2024). A sociocultural approach to voting: Construing voting as a duty to others predicts political interest and engagement. Proceedings of the National Academy of Sciences, 121(22), e2215051121. https://doi.org/10.1073/pnas.2215051121
\\[0.5\baselineskip]
Wulff, D. U., \& Mata, R. (2025). Semantic embeddings reveal and address taxonomic incommensurability in psychological measurement. Nature Human Behaviour, 9(5), 944--954. https://doi.org/10.1038/s41562-024-02089-y
\\[0.5\baselineskip]
Zhao, Y., Qiao, T., Chen, Y., Kuang, M., Bai, W., Yi, Y., Huang, X., Li, W., \& Wang, W. (2025). Attention on social media depends more on how you express yourself than on who you are. Nature Human Behaviour, 10(2), 288--302. https://doi.org/10.1038/s41562-025-02323-1
\\[0.5\baselineskip]
Zöller, N., Berger, J., Lin, I., Fu, N., Komarneni, J., Barabucci, G., Laskowski, K., Shia, V., Harack, B., Chu, E. A., Trianni, V., Kurvers, R. H. J. M., \& Herzog, S. M. (2025). Human--AI collectives most accurately diagnose clinical vignettes. Proceedings of the National Academy of Sciences, 122(24), e2426153122. \\https://doi.org/10.1073/pnas.2426153122
\end{quote}

\subsection*{Supplementary text}\label{appendix-b-additional-results}

\noindent\textbf{Details on model and decoding strategy selection and justifications}

In addition to the variation in prompt, procedure for extracting
quantitative answers from model output, and handling of non compliant
responses discussed in \S \ref{sec:findings-operational}, we also found similar variation in the choice of model and decoding function among papers and tasks in our corpus.

Researchers most commonly used OpenAI models, but among OpenAI models we observed a great deal of variation in model, version, and API source
(e.g. OpenAI vs. Azure). In addition, we observed the use of many other
proprietary and open models. For commercial models, a high-level product
name (i.e. ``ChatGPT'') was often reported rather than a specific model
version (i.e. ``gpt-4o-2024-05-13''), though these high-level product
names are not sufficient for replicability
~\cite{barrieReplicationLanguageModels}. Researchers mostly use a model's reputation or recent research
to justify selecting it, for example choosing a model because it is
``the most widely used and often best-performing LLM available''
~\cite{heseltinePartisanTemporalSelective2025} or because of ``recent work showing that {[}the model{]} can be
used to reliably code text''
~\cite{waldfogelSocioculturalApproachVoting2024a}. Like the justifications we found for prompt design, these are
reasonable explanations given the limited transparency about model
capabilities and biases. However, these justifications are fundamentally
limited: a model\textquotesingle s reputation or success on one task
offers little guarantee that it is ideal for another measurement task.
Instead, different models may perform better on certain tasks and may
exhibit different political, social, and economic biases, all of which
can impact their performance in social science research
~\cite{feng-etal-2023-pretraining, santurkarWhoseOpinionsLanguage2023}.

Likewise, decoding strategies, which refer to how words are selected
during generation, can greatly impact content, repetitiveness,
coherence, and accuracy of the outputs of LLMs
~\cite{dillionCanAILanguage2023}, and may therefore affect the validity of the measurements
produced. We found that most researchers set the temperature to zero to
maximize reproducibility.\footnote{The decoding function controls the
  randomness of model outputs; for instance, setting temperature to zero
  produces deterministic (minimally random) outputs, while higher
  temperatures increase variability.} Several papers instead use
decoding strategies that maximize randomness of generated responses
(e.g. set the temperature to 1) and average over multiple model outputs,
either to produce multiple ``independent'' annotations
\cite{hurLanguageSentimentPredicts2024a},
\cite{lemensUncoveringSemanticsConcepts2023}
or to try to mimic the natural variation in human survey responses
\cite{bisbeeSyntheticReplacementsHuman2024}.

\vspace{2\baselineskip}
\setcounter{table}{0}
\renewcommand{\thetable}{S\arabic{table}}

\noindent\textbf{Table S1: Counts of selected papers and tasks which met our inclusion criteria across journals.}

\begin{table}[h!] \label{tab:journals}
\centering
\begin{tabular}{lrrrr}
\hline
\textbf{Journal} 
  & \textbf{\# Total} 
  & \textbf{\# Keyword} 
  & \textbf{\# Selected} 
  & \textbf{\# Selected} \\
  & \textbf{papers} 
  & \textbf{selected} 
  & \textbf{articles} 
  & \textbf{tasks} \\
\hline
PNAS                                          & 568  & 97  & 13 & 28 \\
Nature Human Behaviour                        & 539  & 97  &  7 & 11 \\
Political Analysis                            & 144  & 26  &  4 &  8 \\
American Political Science Review             & 452  & 17  &  2 &  2 \\
American Journal of Political Science         & 440  & 10  &  1 &  1 \\
American Journal of Sociology*                & 138  & 12  &  0 &  0 \\
Annual Review of Sociology*                   & 100  & 18  &  0 &  0 \\
American Sociological Review*                 & 144  &  9  &  0 &  0 \\
\hline
\textbf{Total} & \textbf{2143} & \textbf{246} & \textbf{27} & \textbf{50} \\
\hline
\end{tabular}

\end{table}
 {\small * Rather than screening bulk-downloaded articles as for the other venues, we searched these journals through their online search interfaces using the same keywords. No returned articles met our inclusion criteria.}

\newpage

\noindent\textbf{Table S2: Overview of use-cases of LLM-generated measurements.}
In most cases, LLM-generated measurements
play a central role in a paper's core empirical claim.

\begin{table}[h!]
\centering
\begin{tabular}{lrrl}
\hline
\textbf{Purpose of LLM-generated measurements} & \textbf{\# Papers} & \textbf{\# Tasks} & \textbf{Centrality to a} \\
                                                &                    &                   & \textbf{paper's claim}   \\
\hline
Evaluation for social science task                 &  7 & 21 & High   \\
Part of main analysis for social science question  & 12 & 17 & High   \\
Used to validate alternative method in main study  &  7 &  7 & Medium \\
Data selection or filtering                        &  4 &  3 & Medium \\
Exploratory data analysis                          &  1 &  1 & Low    \\
\hline
\textbf{Total*} & \textbf{27} & \textbf{50} & \\
\hline
\end{tabular}
\label{tab:purposes}
\end{table}
{\small * Some papers contain multiple use cases across tasks, hence the Papers column does not sum to the total number of papers.}
\\ \\
\noindent\textbf{Table S3: Reporting practices across papers in our corpus.} Most papers
document the prompt and model used, but fewer report technical details
such as the decoding function, output extraction procedure, or handling
of non-compliant responses.
In most cases, LLM-generated measurements
play a central role in a paper's core empirical claim.

\begin{table}[h!]
\centering
\begin{tabular}{lrr}
\hline
                                                                       & \textbf{\# Papers} & \textbf{\# Tasks} \\
\hline
Prompt reported                                                        & 22 & 45 \\
Model reported                                                         & 26 & 48 \\
Reported decoding function                                             & 11 & 26 \\
Reported how quantitative answer is extracted from the model's output  &  4 &  6 \\
Reported how non-compliant answers are addressed                       & 10 & 21 \\
\hline
\textbf{Total}                                                         & \textbf{27} & \textbf{50} \\
\hline
\end{tabular}

\label{tab:reporting}
\end{table}

\end{document}